# An Integrated Optical Circuit Architecture for Inverse-Designed Silicon Photonic Components

Richard Soref, *Life Fellow, IEEE*, and Dusan Gostimirovic

*Abstract*—In this work, we demonstrate a compact toolkit of inverse-designed, topologically optimized silicon photonic devices that are arranged in a "plug-and-play" fashion to realize many different photonic integrated circuits, both passive and active, each with a small footprint. The silicon-on-insulator 1550-nm toolkit contains a 2 x 2 3-dB splitter/combiner, a 2 x 2 waveguide crossover, and a 2 x 2 all-forward add–drop resonator. The resonator can become a 2 x 2 electro-optical crossbar switch by means of the thermo-optical effect, phase-change cladding, or free-carrier injection. For each of the ten circuits demonstrated in this work, the toolkit of photonic devices enables the compact circuit to achieve low insertion loss and low crosstalk. By adopting the sophisticated inverse-design approach, the design structure, shape, and sizing of each individual device can be made more flexible to better suit the architecture of the greater circuit. For a compact architecture, we present a unified, parallel-waveguide circuit framework into which the devices are designed to fit seamlessly, thus enabling low-complexity circuit design.

## I. INTRODUCTION

SILICON photonics has experienced remarkable growth in cost-effective applications including optical computing [1–3], sensing [4,5], and optical communications [6–8]. The silicon-photonics platform leverages the modern CMOS foundry-fabrication infrastructure to yield fully integrated electronic-photonic wafers and chips. Photonic inverse design has been introduced recently as a method of accelerating and automating the design of advanced chips [9,10].

Silicon-on-insulator (SOI) photonic integrated circuits (PICs) are constituted of waveguided active and passive components. Some components have become ubiquitous, such as directional couplers, waveguide crossovers, and microring resonators (MRRs) that are side coupled to bus waveguides. Others, important but less pervasive, are star couplers, mode converters, N x M multimode interferometers, 1 x N splitters, and N x 1 combiners. Generally, all of these can be created using the inverse-design approach. Each of the components have issues of large footprint or large optical crosstalk, or, for the MRR, the second bus feeds backward instead of forward. Such issues "motivate" the inverse-designed component (IDC) approach. The IDC approach investigated here can resolve many of these issues, but at the expense of increased fabrication complexity. To create IDCs, the high-resolution photolithography of the foundry is required to define a pixelated array of tiny air holes (later filled with oxide) within an initially uniform ~220 nm silicon rectangular film that connects input strip waveguides to the output strip waveguides. This photoetching of silicon is done in one processing step for all IDCs in the circuit. We speculate that the tradeoff in complexity will be worthwhile because of the enhanced component performances.

The investigation presented in this work focuses on three IDCs, each having two input ports and two output ports: a 3 dB coupler (splitter/combiner), a 100% coupler (crossover), and a new square micro-resonator with feed-forward behavior. Electro-optical crossbar switching of the square resonator (also known as reconfiguration of an add–drop multiplexer) is simulated here by means of a uniform change in the real index of the pixelated silicon film, accomplished, for example, by the thermo-optical (TO) effect. Examples of these IDCs working together cooperatively in PICs are presented. This paper focuses on the SOI platform operating at the 1550 nm telecom wavelength; however, our IDC approach applies to any "on-insulator" semiconductor platform, which means any group IV, III-V, or II-VI semiconductor circuit on the oxidized silicon wafer-substrate. In the context of wavelength-division multiplexed on-chip systems, we present passive and active (switched) IDC circuits. Ten specific circuit applications are illustrated in this paper.

## II. THE INVERSE-DESIGN APPROACH

The literature reports IDC results in coarse wavelength division (de)multiplexers (CWDM) [11,12], mode converters [13,14], bends [15], and waveguide crossings [16,17] with performance per unit area far greater than that of conventional, hand-based designs. Such design is initiated by the desired objective function (e.g., maximizing modal throughput from one input waveguide to another output waveguide) and by defining a highly dimensional design space (often in the form of a "pixelated" permittivity matrix) to be modified by the inverse-design algorithm. Perhaps the most successful of these

This paragraph of the first footnote will contain the date on which you submitted your paper for review. It will also contain support information, including sponsor and financial support acknowledgment. For example, "This work was supported in part by the U.S. Department of Commerce under Grant BS123456."

Richard Soref is with the Engineering Department, University of Massachusetts, Boston, MA 02125 USA (e-mail: richard.soref@umb.edu).

Dusan Gostimirovic is with the Department of Electrical and Computer Engineering, McGill University, Montreal, QC H3A 0G4 Canada (e-mail: dusan.gostimirovic@mcgill.ca).

inverse design algorithms is topology optimization [18,19], which iteratively defines the hundreds-to-thousands of material permittivity pixels in the design area to maximize the specified objective function—producing highly nonintuitive designs that are highly unlikely to be achieved by hand design. Optimization of such a highly dimensional design problem is made possible by the adjoint method [20,21], which requires only two simulations to be made per optimization iteration, where the design variables (permittivity pixels) are ultimately updated based on the gradient of their effect on the objective function.

For system-level design, where individual components may have to share key parameters and fabrication process-specific design constraints, inverse design is a useful method, as the skeleton of the framework can be generally defined by the designer, and the blanks (the device design areas) can be filled in to suit the desired functionality, performance targets, and design constraints. In this work, we present a unified framework for easily arrangeable photonic circuits built with inverse-designed components. This framework is a series of parallel waveguides with fixed spacing, but a variable number of waveguides depending on the function and scale of the circuit. For example, a monochromatic 8 x 8 crosspoint matrix switch uses 16 parallel waveguides, and a monochromatic 8 x 8 Spanke–Benes permutation matrix switch uses eight. Having set the framework, a set of "plug-and-play" building block devices can be arranged within it to carry out the desired functionality.

In this work, we present three inverse-designed, topologically optimized SOI photonic building blocks: the 2 x 2 3-dB splitter/combiner, the 2 x 2 waveguide crossover, and the 2 x 2 *all-forward* add–drop resonator. Each of these devices feature two parallel waveguides with the same size and spacing to match the circuit framework, and a topologically optimized design region within the center to carry out the optical functionality. To demonstrate the flexibility of this approach, we present the designs of ten different circuits.

We aim to create a simplified framework for PICs to promote design uniformity, short design times, low complexity, low spatial footprint, high optical performance, and high signal quality. To achieve these goals, the core components (devices) use the highest quality design methods, and the unifying framework must be set up to support these devices with a structure that suits them best. We believe the method of topological inverse design produces the best performing devices; given the parallel input/output structure generally used with the algorithm, we believe a parallel series of waveguides best supports them while keeping circuit complexity low. The design of the circuit and each individual component follow the same "fill-in-the-blank" process: for the devices that have a high degree of design dimensionality, we delegate the design work to a computer algorithm; for the circuit, which has a low degree of dimensionality, we do the design work by hand.

## III. THE INVERSE-DESIGNED COMPONENTS

The 2 x 2 3-dB splitter/combiner, 2 x 2 crossover, and 2 x 2 all-forward add–drop resonator presented in this section are designed with topology optimization using the adjoint method from the open-source topology optimization software, Angler [21], where a 2.5D effective-index FDFD solver is used for device simulation. For this work, modifications were made to the software to enable multi-objective optimizations. For the first two devices, we initially considered two parallel waveguides spaced 2 μm apart (forming two input waveguides and two output waveguides) and an inverse design region in the center of the two parallel waveguides. This would demonstrate small footprint while achieving high optical performance. The required size and waveguide spacing is larger for the resonator, however, and so, for consistent integration of components, we then created a new splitter/combiner and a new crossover whose spacings matched that of the resonator (10 x 10 μm).

The optimizations are set up for each device to maximize optical throughput for a given mode and given wavelength for each initial condition of each device's targeted functionality. The inverse design region of each device is discretized into a matrix of 20-nm pixels that can take a permittivity value of silicon's ($\varepsilon_{r,Si} = 12.1$) or silica's ($\varepsilon_{r,SiO2} = 2.07$) based on the direction the optimizer takes in maximizing the objective function. To efficiently optimize the hundreds of pixels in the design region, topology optimization makes use of the adjoint method to calculate the optimization gradient, which only requires two simulations to be made per iteration. The first simulation is a typical simulation of the device; the second simulation swaps the input and output directions to make a "reverse simulation" to find the gradient of the objective function with respect to the set of design parameters (pixel permittivity values). This method has shown success in designing compact high-performance silicon-photonic devices and is easily adaptable to new devices with unique design elements and functionalities, such as ours.

### A. 2 x 2 3-dB splitter/combiner

The 2 x 2 3-dB splitter/combiner, as shown in Fig. 1 is the essential passive component of Mach–Zehnder interferometer switches. Its functionality is such that a signal at either of the two input waveguides is split with a 50:50 ratio at the two output waveguides, ideally with minimal loss and back reflection. Should a same-wavelength signal be present at each input, the two combine at one of the two output ports, depending on if the two signals are in phase or out of phase. The device features the standard, 220-nm SOI platform, with 500-nm-wide input- and output waveguides spaced 9 μm apart. A dimension of 10 x 10 μm² is chosen for the inverse design region (square component in the middle of the device) to match the other two devices presented later in this work.

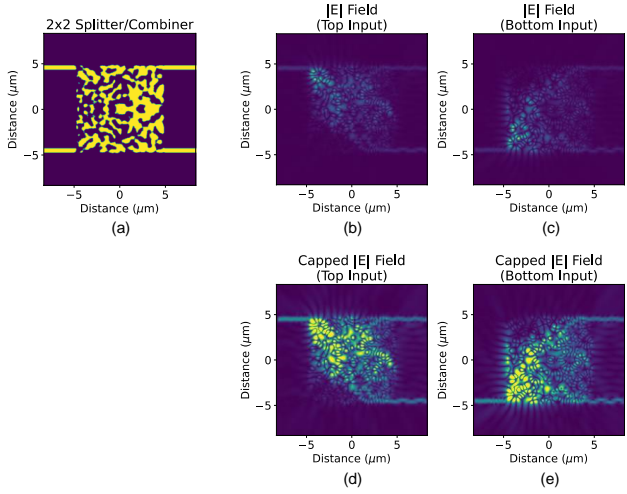

Fig. 1. (a) Topologically optimized 2 x 2 3-dB splitter/combiner (in splitter mode) and the top-view optical field profiles for the following initial conditions: (b) input at the top waveguide, (c) input at the bottom waveguide. Capped field values of (b) and (c) are shown in (d) and (e), respectively, to better visualize the field path through the design region.

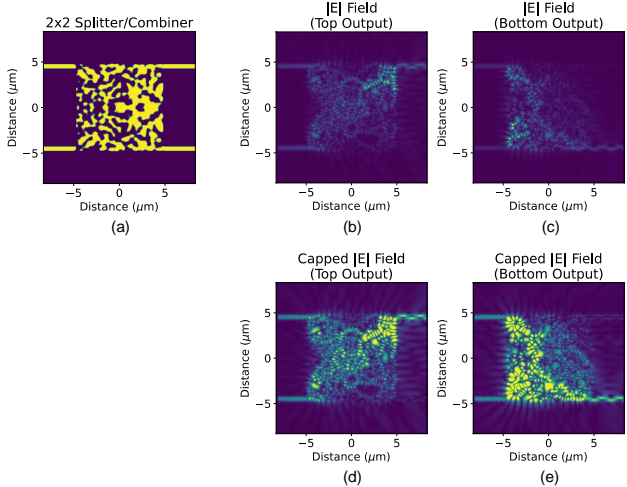

Fig. 2. (a) Topologically optimized 2 x 2 3-dB splitter/combiner (in combiner mode) and the top-view optical field profiles for the following initial conditions: (b) inputs at both waveguides with $\pi/2$ phase difference, and (c) inputs at both waveguides with $-\pi/2$ phase difference. Capped field values of (b) and (c) are shown in (d) and (e), respectively, to better visualize the field path through the design region.

The topological optimization is set up to maximize 50:50 splitting for a 1550-nm $TE_0$ mode, whether it comes in at the top input waveguide or the bottom input waveguide. The combiner functionality is not explicitly included in the optimization, as we can expect complete reciprocity of the splitter function in this passive optical device. A large spatial filter (commonly applied to modern topologically optimized devices [22]) is applied to finalize a design that contains typically larger features that are fabricated more reliably. In general terms, a smaller spatial filter produces designs with smaller, more difficult to fabricate features, but also allows for higher performance, as a larger design space can be explored in optimization. We carefully manage this tradeoff to avoid extremes.

Figure 1 also shows the top-view optical field profiles of the device for each expected initial condition (other than no input being present). A 50:50 (top:bottom) split is visualized for the first two initial conditions. The exact splitting ratios at 1550 nm are found to be 0.466:0.490 and 0.456:0.472 for top-waveguide and bottom-waveguide inputs, respectively, which indicate low insertion loss (0.20 dB and 0.32 dB). For the two initial conditions that satisfy the combiner function, as shown in Fig. 2, the phase of the two inputs determine which output port the signals combine at. For phase conditions A (inputs have a $\pi/2$ phase difference) and B (inputs have a $-\pi/2$ phase difference), the combining ratios are 1.840:0.042 and 0.003:1.882, respectively, which indicate low insertion loss (0.36 dB and 0.26 dB) and low crosstalk (-16.8 dB and -28.2 dB

Like that of the nonintuitive inverse design region, the optical field profile within follows a pattern that is difficult to decode. We see that the signals take a nonlinear path from the input to the output, making use of the entire inverse design region, but remain well-confined within it. Given the extremely high dimensionality of the optimization problem, it is unsurprising that the performance and complexity of the final device design region are both high.

Using two Fig.-1 devices, there is a simple IDC procedure to create a high-performance broad-spectrum 2 x 2 Mach-Zehnder-interferometer cross-bar switch. First, two parallel waveguides are used to connect the two outputs of the first Fig.-1 splitter to the two inputs of the second Fig.-1 combiner. Then, an electro-optical phase shifter (EOPS) is inserted into one of the two connecting waveguide arms. The bar state of the 2 x 2 is a stable non-volatile state where the EOPS is zero. The cross state is reached when EOPS is $\pi$ radians of shift.

### B. 2 x 2 crossover

The 2 x 2 crossover, as shown in Fig. 3, is the second key passive component of higher order integrated optical switches. Its functionality is such that a signal at either of the two input waveguides will output at the diagonally opposite port (cross state), with minimal loss and back reflection. Like the 2 x 2 3-dB splitter/combiner, this device features the standard, 220 nm SOI platform, with 500-nm-wide input and output waveguides spaced 9 μm apart. The same compact dimension of 10 x 10 $\mu m^2$ is chosen for the design region. The two devices presented thus far appear to be similar, but the highly dimensional design regions are distributed differently.

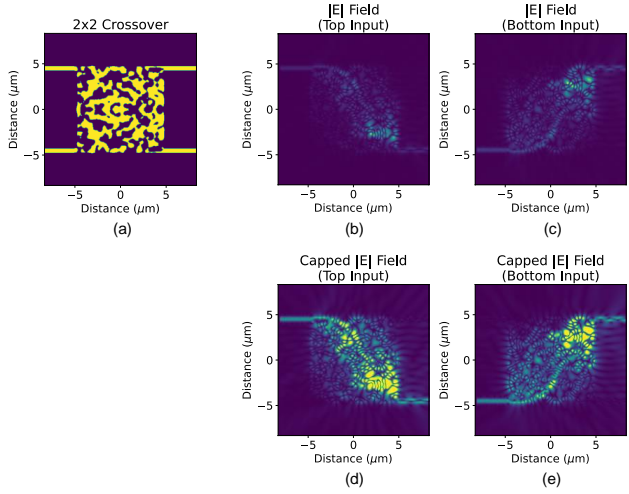

Fig. 3. (a) Topologically optimized 2 x 2 3-dB crossover and the top-view optical field profiles for the following initial conditions: (b) input at the top waveguide and (c) input at the bottom waveguide. Capped field values of (b) and (c) are shown in (d) and (e), respectively, to better visualize the field path through the design region.

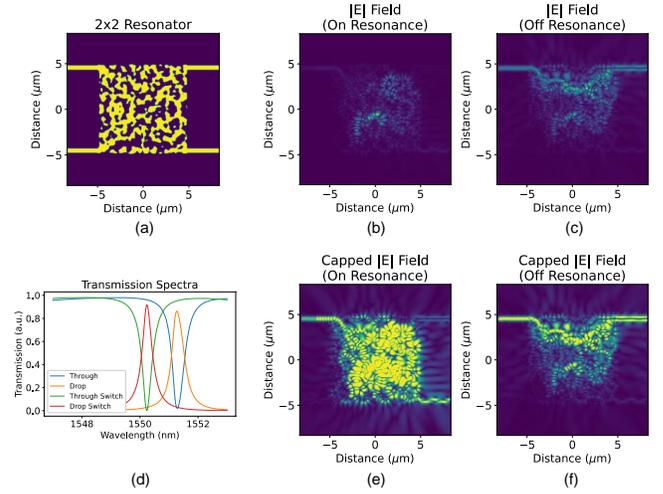

Fig. 4. (a) Topologically optimized all-forward add–drop resonator and (d) the optical transmission spectra at both outputs. The top-view optical field profiles for the following conditions: (b) through-port output (off resonance) and (c) drop-port output (on resonance). Capped field values of (b) and (c) are shown in (e) and (f), respectively, to better visualize the field path through the design region.

The topology optimization is set up to maximize power transfer from the top input waveguide to the bottom output waveguide, and from the bottom input waveguide to the top output waveguide, for a 1550-nm $TE_0$ mode. To maintain similarity with the 2 x 2 3-dB splitter/combiner, the same large spatial filter is also applied here to obtain a final design that contains the same easy-to-fabricate large features. Figure 3 also shows the top-view optical field profiles of the device for each of the two previously mentioned initial conditions, where we observe almost complete waveguide crossing for both. The exact crossing ratio is 0.003:0.959 (top:bottom) and 0.911:0.001 for optical stimulation at the top and bottom input waveguides, respectively, which indicates low insertion loss (0.18 dB and 0.40 dB) and low crosstalk (-25 dB and -29 dB).

### C. 2 x 2 all-forward add–drop resonator

The 2 x 2 all-forward add–drop resonator, as shown in Fig. 4, is the third key passive component intended to be the functional equivalent of the two-bus-coupled MRR, but here a monolithic shape of silicon without evanescent coupling. Like the MRR, an off-resonance signal at the top left input waveguide will exit at the top right output waveguide; but different from the MRR, the resonance signal here exits at the bottom right output instead of the bottom left. To the best of our knowledge, this is the first gap-less all-forward 2 x 2 resonator—a device designed by topology optimization for high levels of performance in a compact spatial footprint.

Like the other two devices in this toolkit, this device features the standard, 220 nm SOI platform, with 500-nm-wide input and output waveguides spaced 9 µm apart (vertically). To achieve good performance (i.e., high extinction ratio and quality factor), the design region is set to 10 x 10 µm². In general terms, topologically optimized devices perform better than their conventionally designed counterparts per the same unit area. Although this device is 2x larger than the "smallest" 2 x 2 MRRs, the all-forward arrangement is a complex constraint for the optimization problem and more room must be given for the optical signal to diverge from its natural "resonant path" and exit in the forward direction.

The topological optimization is set up to maximize power transfer from the top left input waveguide to one of the two right-side output waveguides in such a way that light is maximized at the drop port (bottom output) for 1550 nm and is maximized at the through port (top output) for 1548 nm and 1552 nm (again for a $TE_0$ mode). This objective function guides the optimizer to create a resonator with a large extinction ratio for a quality factor of approximately 4,500, as shown in the Lorentzian-like transmission spectra in Fig. 4d. Like photonic crystal nanocavities, this device features only one peak, which carries benefit in high-throughout, multichannel circuits. A larger quality factor of the peak can be achieved with a tradeoff in optimization time, extinction ratio, or feature size/complexity. Likewise, a smaller quality factor can improve the other metrics. The transmission plot also shows the spectra for the through and drop ports for a shift in real refractive index of the silicon-square of $\Delta n = 0.003$, (a resonance shift of 2 linewidths), which indicates the device's ability to achieve full crossbar switching under electro-thermal control, for example.

Figure 4 also shows the optical field profiles from the top view of the device, for each of the two previously mentioned initial conditions. In Fig. 4e, the resonance-enhanced buildup of light is clearly shown near the middle of the design region, before exiting diagonally at the drop port. For the off-resonance condition shown in Fig. 4f, the optical signal takes a nonintuitive path partially into the design region but gets redirected back to the top, through port. This nonlinear path is likely a cause for the through-port (bar state) insertion loss and can likely be reduced with further tuning of the optimization length and parameters.

## IV. Photonic Switch Layouts

The following ten circuits are examples of integrated photonic switches composed of the same core building blocks presented in the previous section. Each circuit also features the same parallel-waveguide architecture, where individual devices are placed in a "plug-and-play" fashion to realize unique functionalities in a unified design strategy. By (inverse) designing the core building blocks to be compact, with low insertion loss and low crosstalk, these large, complex circuits can be realized in a simple and efficient manor, ultimately bringing large performance improvements to the chip.

### A. Monochromatic 8 x 8 crosspoint matrix switch

The term monochromatic means that the user choses a particular wavelength of operation $\lambda_0$ for all signals. After that, the 2 x 2 resonators are designed so that the resonance wavelength $\lambda_r$ matches $\lambda_0$. We assume here that each resonator is an individually controlled, electrically controlled 2 x 2 switch, and, to achieve switching, we can identify three EO resonance-shifting mechanisms: (1) the TO effect in the silicon design region square, where $\Delta n$ is triggered by a very-nearby nano-scale electrical "Joule heater;" (2) the phase-change-material approach in which a PCM cladding (like $Sb_2Se_3$) upon the silicon-square top surface has its phase changed by nano-heaters; or (3) lateral injection of free electrons and holes into the intrinsic silicon square by forward bias applied across P- and N-doped regions at opposite edges of the square.

The most classical N x N switch is the "crossbar," a matrix comprised of 2 x 2 crosspoints (here, resonant ones), and here in our parallel-waveguide approach, we can attain this classical geometry without using any waveguide crossovers. Our design is presented in Fig. 5, and here we can illustrate the matrix operation for the TO case by saying that all 2 x 2s are initially in the cross state, which is the unheated ambient state. Here the rows and columns are slanted, and we use coincident row–column addressing of one crosspoint in each row to produce a bar state in each row. In Fig 5 and the subsequent figures, the electrical control wires are not shown, for simplicity. The x–y addressing means that, out of the $N^2$ devices, only N are addressed at any one time.

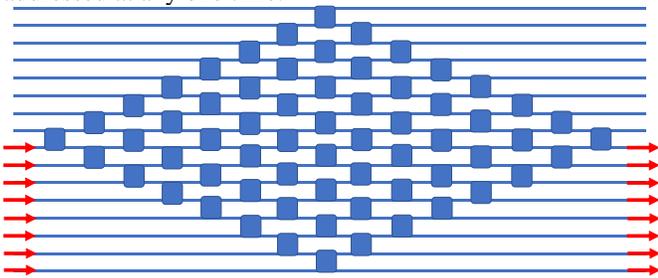

Fig. 5. Block diagram of the monochromatic 8 x 8 crosspoint matrix switch. Each blue square indicates an electrically tunable resonant add–drop IDC.

### B. Monochromatic 8 x 8 Spanke–Benes permutation matrix switch

In the matrix art, it is known that there are architectures for N inputs and N outputs that have fewer switches than does the crossbar. One such example is the Spanke–Benes permutation matrix switch, requiring only 28 elements for 8 x 8 switching, as we show in Fig. 6. Here again, we attain this matrix without using any waveguide crossovers.

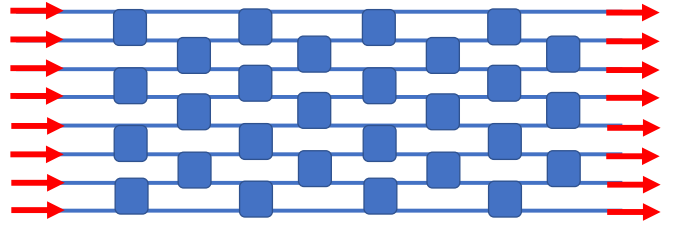

Fig. 6. Block diagram of the monochromatic 8 x 8 Spanke–Benes permutation matrix switch. This diagram is based upon the layout of Fig. 6b in [23].

### C. Monochromatic 8 x 8 PILOSS topology matrix switch

For a large-scale addressed matrix in which light travels through various paths, a path-independent insertion loss is a desirable attribute, and for this constraint, the PILOSS topology has been invented [23]. Our next IDC circuit example is based upon the block diagram given in Fig. 6c of [23], and here 64 resonator switches and 49 crossovers are deployed for the 8 x 8 PILOSS topology, as presented in Fig. 7. Like the other circuits, a uniform parallel array of waveguides is the framework.

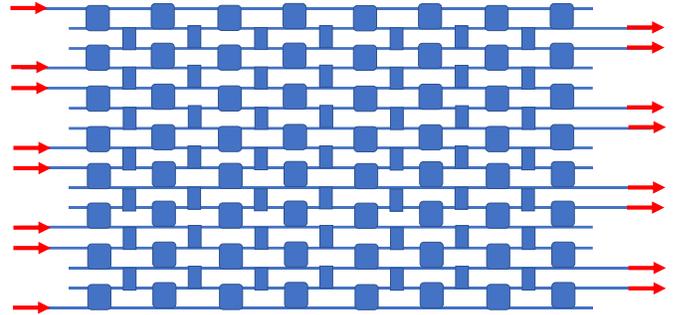

Fig. 7. Block diagram of the monochromatic 8 x 8 PILOSS-topology matrix switch. Each narrow blue rectangle represents a passive IDC crossover, with the squares indicating EO switches. This diagram is based upon the diagram in Fig. 6c of [23].

### D. Monochromatic non-blocking 16 x 16 Clos–Benes spatial routing switch

In the photonics literature, the non-blocking Clos-Benes architecture is an important approach to large-scale routers because the number of elemental switches required is quite low compared to the number in other approaches. However, there is a drawback, which is the "perfect shuffle interconnections" that are required within the matrix, and these consist of a complex array of waveguide crossings at oblique angles. We have invented here a simple triangular array of IDC crossings that perform the needed shuffle, and this method is shown in Fig. 8 for the 16 x 16 routing case. Note that here the number of EO switches is reduced to only 40. A fairly large number of passive crossovers (narrow blocks) is required, as indicated.

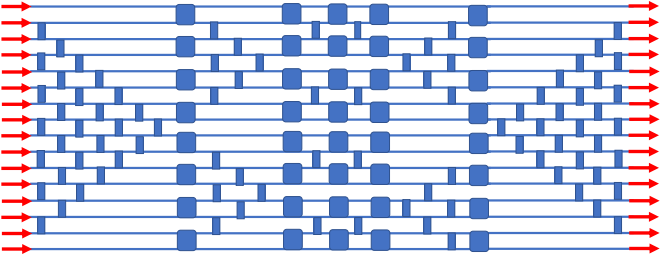

Fig. 8. IDC block diagram of the monochromatic non-blocking 16 x 16 Clos–Benes spatial routing switch. This topology is based on the schematic in Fig. 4 of [24].

### E. Monochromatic 8 x 1 and 1 x 8 spatial routing switch

Some applications require active N x 1 or 1 x N routing, and for this we have a simple approach that is presented in Fig. 9. It is interesting that no crossovers are needed for this N = 8 case. The 8 x 1 concept is to route a selected input while blocking the other seven input signals. The small purple squares in Fig. represent absorbers or waveguide "terminators."

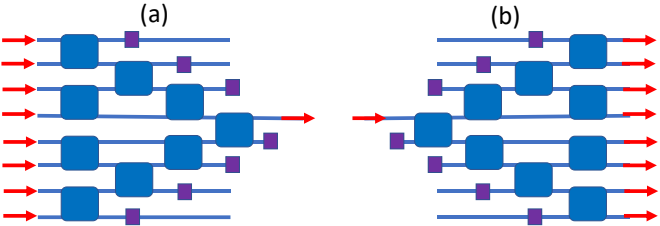

Fig. 9. IDC block diagram on monochromatic 8 x 1 and 1 x 8 spatial routing switches.

### F. 8 x 1 and 1 x 8 passive wavelength-division multiplexer

At this point, we move into the domain of wavelength-multiplexed circuits and networks. Here, we use a group of 2 x 2 resonators, and we require that each resonator is "dedicated" to a particular $\lambda_r$ or "color," with that color being different from all other colors in the group. In our diagrams, we "paint" a color on each passive add–drop resonator to indicate its $\lambda_r$, such as red, orange, green, blue, violet, navy, or sienna. Having done that, we then include a group of broadband 2 x 2 crossovers shown as narrow blue rectangles, where the 28-nm wavelength-bandwidth of each crossover encompasses all the colors. The new design is presented in Fig. 10 for MUX and DEMUX. In this design, it is assumed that an add–drop is in cross state for its color and is in the bar state for any other color. For this MUX/DEMUX, the add–drops are passive.

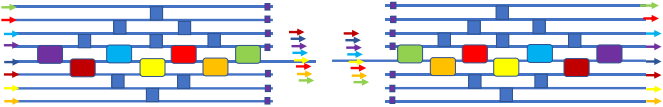

Fig. 10. Block diagrams of the wavelength-division (a) 8 x 1 multiplexer and (b) 1 x 8 demultiplexer.

### G. 2 x 2 x 3λ multi-crossbar switch

Now let us add EO resonance shifting (electrical control) to each wavelength-dedicated 2 x 2 add–drop. A simple-but-effective switch is the cascade connection of three different 2 x 2 switches, as is shown in Fig. 11a. This is a wavelength-multiplexed crossbar switch, with each 2 x 2 color "element" being independently controlled. The lower diagram shows the composite spectrum of the device, illustrating individual TO shifts for red, green, and blue. As discussed in [25], a given color input is on resonance for one element (the cross state there) and is off resonance for all other elements (the bar state there). A given element can be shifted from the cross state to the bar state. This multi-crossbar approach is illustrated in Figs. 3 and 4 of [25].

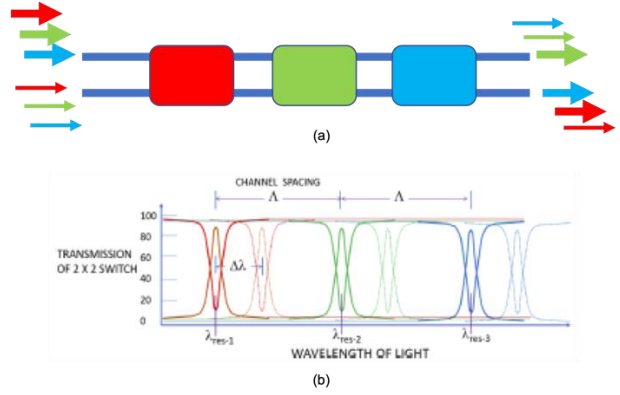

Fig. 11. Block diagram of (a) the 2 x 2 x 3λ multicrossbar switch in a series connection of three resonant devices, and (b) the overall spectral response for the two states of each constituent.

### H. 6 x 6 x 4λ wavelength-selective switch

Having just described the operation of the 2 x 2 x Nλ crossbar, let us now put these devices to work in a larger-scale M x M x Nλ wavelength-selective routing switch for the case of 6 x 6 x 4λ, which means that that there are four "color signals" at each of the six inputs and re-arranged (switched) four color signals at each of the four outputs. Our new design is given in Fig. 12, where we also deploy multiple broadband 2 x 2 crossovers (blue squares here). This diagram is a highly modified version of the monochromatic 6 x 6 utilizing MRRs presented in [26], where one modification is the replacement of MRRs with our four-color crossbars. If we visualize this as a 24 x 24 router, then the number of active and passive components employed to accomplish the wavelength routing (48 and 9, respectively) is relatively small.

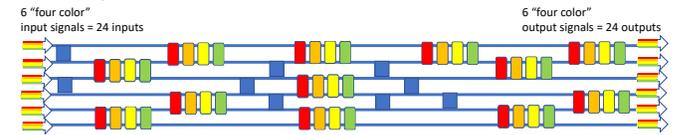

Fig. 12. Block diagram of the 6 x 6 x 4λ wavelength-selective switch in 6 stages.

### I. 8 x 8 x 3λ wavelength-selective switch

Our next WSS has eight input channels and eight output channels, each containing three "color signals." Starting from the monochromatic 8 x 8 in Fig. 5 of [27], we expanded that to the wavelength-MUXed application by employing our above Fig.-11 device to invent a "24 x 24" router that employs only 60 active devices. This is shown in Fig. 13.

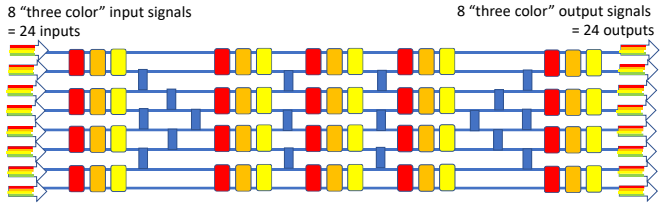

8 "three color" input signals = 24 inputs

8 "three color" output signals = 24 outputs

Fig. 13.   Block diagram of the 8 x 8 x $3\lambda$ wavelength-selective switch.

### J. 4 x 4 x $4\lambda$ wavelength cross-connect switch

The MRR-based "switch and select" approach for wavelength cross-connect switching was recently explored [28], but that approach demands a complicated "shuffle-like" waveguide interconnection. Here we have embodied the switch-and-select architecture with IDCs, not MRRs, for our last EO PIC proposal. However, we shall deviate from the assumption of an everywhere-parallel waveguide framework because the multi-oblique interconnection cannot be readily attained with IDC crossover arrays, and so the 4 x 4 x $4\lambda$ cross-connect switch illustrated here is a "hybrid" that includes the parallel framework and well as a "multi-crossed waveguides" region. Figure 14 presents this WCC which has 48 active devices and 16 passive devices to route 12 incoming WDM signals to 12 outgoing WDM signals. In the parallel region, no crossovers are needed. This nonblocking switch offers low crosstalk.

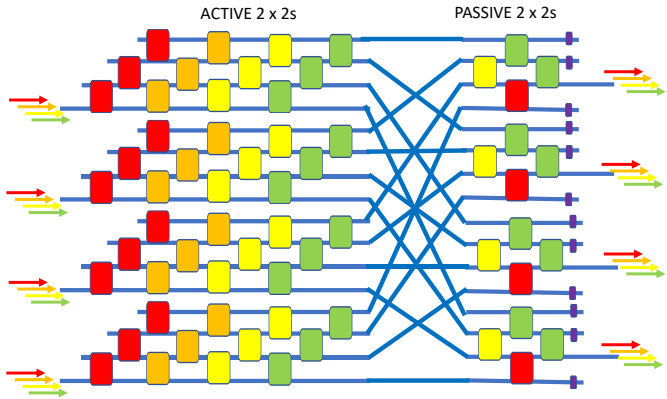

ACTIVE 2 x 2s            PASSIVE 2 x 2s

Fig. 14.   Block diagram of the 4 x 4 x $4\lambda$ wavelength cross-connect switch.

## V. Conclusion

We have deployed a unified design strategy to create a compact collection of inverse-designed, topologically optimized, integrated silicon-photonic devices: a 10 x 10 μm² 2 x 2 3-dB splitter/combiner, a matching 2 x 2 crossover, and a matching 2 x 2 all-forward add–drop resonator. The passive add–drop resonator can be converted into an electrically controlled 2 x 2 crossbar switch, where the proposed EO mechanisms are the TO effect, the phase-change-cladding effect (actuated via a nano-heater), or free-carrier injection via lateral PIN structure. The unified shape, input–output positioning, and sizing allow these IDCs to be easily arranged in a simplified, parallel-waveguide circuit architecture to perform many different switching and computational functionalities—of which we demonstrate ten different examples in this work in the context of WDM chips. Because of the high efficiency of topological optimization with the adjoint method, extremely high-dimensional design spaces can be explored to create highly performing devices within constricted spatial footprints, such as ours. In other words, by leveraging the power of photonic inverse design, the complexity of circuit design can be significantly reduced while also improving performance. Low levels of insertion loss and of crosstalk are achieved for each of the three devices presented from our toolkit, which ultimately enables larger-scale, lower-power, and higher-bandwidth PICs for next-generation communications and computing applications.

## Acknowledgment

The work of RS is supported by the Air Force Office of Scientific Research on grant FA9550-21-1-0347.